\documentclass[12pt]{article}
\usepackage{graphicx}
\usepackage{naturenew}
\usepackage{naturefem}
\usepackage{citesupernumber}
\usepackage{nature}
\usepackage{setspace}

\begin{document}

\bibliographystyle{nature}

\title{Neutral Theory and Relative Species Abundance in Ecology}

\author{Igor Volkov$^1$, Jayanth R. Banavar$^1$, Stephen P. Hubbell$^2$
and \\ Amos Maritan$^3$}

\date{}

\maketitle

1 Department of Physics, 104 Davey Laboratory, The Pennsylvania State University, University Park, Pennsylvania
16802 USA

2 Department of Plant Biology, The University of Georgia, Athens, GA 30602 USA and The Smithsonian Tropical
Research Institute, Box 2072, Balboa, Panama

3 International School for Advanced Studies (S.I.S.S.A.), Via Beirut 2/4, 34014 Trieste, INFM and The Abdus
Salam International Center for Theoretical Physics, Trieste, Italy

\begin{abstract}

The theory of island biogeography\cite{MacArthur1} asserts that an
island or a local community approaches an equilibrium species
richness as a result of the interplay between the immigration of
species from the much larger metacommunity source area and local
extinction of species on the island (local community).
Hubbell\cite{Hubbell1} generalized this neutral theory to explore
the expected steady-state distribution of relative species
abundance (RSA) in the local community under restricted
immigration. Here we present a theoretical framework for the
unified neutral theory of biodiversity\cite{Hubbell1} and an
analytical solution for the distribution of the RSA both in the
metacommunity (Fisher's logseries) and in the local community,
where there are fewer rare species. Rare species are more
extinction-prone, and once they go locally extinct, they take
longer to re-immigrate than do common species. Contrary to recent
assertions\cite{McGill1}, we show that the analytical solution
provides a better fit, with fewer free parameters, to the RSA
distribution of tree species  on Barro Colorado Island
(BCI)\cite{Condit1} than the lognormal
distribution\cite{Preston1,May1}.

\end{abstract}

The neutral theory in ecology\cite{Hubbell1,Bell1} seeks to
capture the influence of speciation, extinction, dispersal, and
ecological drift on the  RSA under the assumption that all species
are demographically alike on a per capita basis. This assumption,
while only an appro\-xi\-ma\-tion\cite{Diamond1,Tilman1,Weiher1},
appears to provide a useful description of an ecological community
on some spatial and temporal scales\cite{Hubbell1,Bell1}. More
significantly, it allows the development of a tractable null
theory for testing hypotheses about community assembly rules.
However, until now, there has been no analytical derivation of the
expected equilibrium distribution of RSA in the local community,
and fits to the theory have required simulations\cite{Hubbell1}
with associated problems of convergence times, unspecified
stopping rules, and precision\cite{McGill1}.

The dynamics of the population of a given species is governed by generalized birth and death events (including
speciation, immigration and emigration). Let $b_{n,k}$ and $d_{n,k}$ represent the probabilities of birth and
death, respectively, in the $k$-th species with $n$ individuals with $b_{-1,k}=d_{0,k}=0$.  Let $p_{n,k}(t)$
denote the probability that the $k$-th species contains $n$ individuals at time $t$. In the simplest scenario,
the time evolution of $p_{n,k}(t)$ is regulated by the master equation\cite{Boswell1,Caraco1,Feller1}:

\begin{equation}
\label{e1} {d p_{n,k}(t)\over d t}=p_{n+1,k}(t)d_{n+1,k}+p_{n-1,k}(t)b_{n-1,k}-p_{n,k}(t)(b_{n,k}+d_{n,k})
\end{equation}
which leads to the steady-state or equilibrium solution, denoted by $P$:

\begin{equation}
\label{e2} P_{n,k}=P_{0,k}\prod_{i=0}^{n-1}{b_{i,k}\over d_{i+1,k}},
\end{equation}
for $n>0$ and where $P_{0,k}$ can be deduced from the
normalization condition $\sum_n P_{n,k}=1$. Note that there is no
requirement of conservation of community size. One can show that
the system is guaranteed to reach the stationary solution
(\ref{e2}) in the infinite time limit\cite{VanKampen1}.

The frequency of species containing $n$ individuals is given by

\begin{equation}
\label{e21} \phi_n=\sum_{k=1}^S I_k,
\end{equation}
where $S$ is the total number of species and the indicator $I_k$ is a random variable which takes the value $1$
with probability $P_{n,k}$ and $0$ with probability $(1-P_{n,k})$. Thus the average number of species containing
$n$ individuals is given by

\begin{equation}
\label{n1} \langle\phi_n\rangle =\sum_{k=1}^S P_{n,k}   .
\end{equation}
The RSA relationship we seek to derive is the dependence of $\langle\phi_n\rangle$ on $n$.

Let a community consist of species with $b_{n,k}\equiv b_n$ and $d_{n,k}\equiv d_n$ being independent of $k$
(the species are assumed to be demographically identical). From Eq.(\ref{n1}), it follows that
$\langle\phi_n\rangle$ is simply proportional to $P_n$, leading to

\begin{equation}
\label{n100} \langle\phi_n\rangle=S P_0\prod_{i=0}^{n-1}{b_i\over d_{i+1}}.
\end{equation}

We consider a metacommunity in which the probability $d$ that an
individual dies and the probability $b$ that an individual gives
birth to an offspring are independent of the population of the
species to which it belongs (density independent case), i.e.
$b_n=b n$ and $d_n=d n$ ($n>0$). Speciation may be introduced by
ascribing a non-zero probability of the appearance of an
individual of a new species, i.e. $b_0=\nu\ne 0$. Substituting the
expressions into Eq.(\ref{n100}), one obtains the celebrated
Fisher logseries\cite{Fisher1} :

\begin{equation}
\label{e5} \langle\phi_n^M\rangle=S_M P_0 {b_0 b_1 ...
b_{n-1}\over d_1 d_2 ... d_n}=\theta {x^n\over n},
\end{equation}
where $M$ refers to the metacommunity, $x=b/d$ and $\theta=S_M P_0
\nu/ b$ is the biodiversity parameter (also called Fisher's
$\alpha$). We follow the notation of Hubbell\cite{Hubbell1} in
this paper. Note that $x$ represents the ratio of effective per
capita birth rate to the death rate arising from a variety of
causes such as birth, death, immigration and emigration. Note that
in the absence of speciation, $b_0 = \nu = \theta = 0$, and, in
equilibrium, there are no individuals in the metacommunity.  When
one introduces speciation, $x$ has to be less than $1$ to maintain
a finite metacommunity size $J_M=\sum_n
n\langle\phi_n\rangle={\theta x\over 1-x}$.

We turn now to the case of a local community of size $J$ undergoing births and deaths accompanied by a steady
immigration of individuals from the surrounding metacommunity. When the local community is semi-isolated from
the metacommunity, one may introduce an immigration rate $m$, which is the probability of immigration from the
metacommunity to the local community. For constant $m$ (independent of species), immigrants belonging to the
more abundant species in the metacommunity will arrive in the local community more frequently than those of
rarer species.

Our central result (see Box 1 for a derivation) is an analytic
expression for the RSA of the local community:

\begin{equation}
\label{e91} \langle\phi_n\rangle=\theta{J!\over n!(J-n)!}{\Gamma(\gamma)\over\Gamma(J+\gamma)}\int_0^\gamma
{\Gamma(n+y)\over\Gamma(1+y)} {\Gamma(J-n+\gamma-y)\over\Gamma(\gamma-y)}\exp(-y \theta/\gamma) d y,
\end{equation}
where $\Gamma(z) = \int_0^\infty t^{z-1} e^{-t} dt$ which is equal to $(z-1)!$ for integer $z$ and
$\gamma={m(J-1)\over 1-m}$. As expected, $\langle\phi_n\rangle$ is zero when $n$ exceeds $J$. The computer
calculations in Hubbell's book\cite{Hubbell1} as well as those more recently carried out by McGill\cite{McGill1}
were aimed at estimating $\langle\phi_n\rangle$ by simulating the processes of birth, death and immigration.

One can evaluate the integral in Eq.(\ref{e91}) numerically for a
given set of parameters: $J$, $\theta$ and $m$. For large values
of $n$, the integral can be evaluated very accurately and
efficiently using the method of steepest descent\cite{Morse1}. Any
given RSA data set contains information about the local community
size, $J$, and the total number of species in the local community,
$S_L = \sum_{k=1}^J \langle\phi_k\rangle$. Thus there is just one
free fitting parameter at one's disposal.

McGill asserted\cite{McGill1} that the lognormal distribution is a
"more parsimonious" null hypothesis than the neutral theory, a
suggestion which is not borne out by our reanalysis of the BCI
data. We focus only on the BCI data set because, as pointed out by
McGill\cite{McGill1}, the North American Breeding Bird Survey data
are not as exhaustively sampled as the BCI data set, resulting in
fewer individuals and species in any given year in a given
location. Furthermore, McGill's analysis seems to rely on adding
the bird counts of $5$ years at the same sampling locations even
though these data sets are not independent.

Figure 1 shows a Preston-like binning\cite{Preston1} of the BCI
data\cite{Condit1} and the fit of our analytic expression with one
free parameter ($11$ degrees of freedom) along with a lognormal
having three free parameters ($9$ degrees of freedom). Standard
chi-square analysis\cite{Press1} yields values of $\chi^2=3.20$
for the neutral theory and $3.89$ for the lognormal. The
probabilities of such good agreement arising by chance are
$1.23\%$ and $8.14\%$ for the neutral theory and lognormal fits,
respectively. Thus one obtains a better fit of the data with the
analytical solution to the neutral theory to BCI than with the
lognormal, even though there are two fewer free parameters.
McGill's analysis\cite{McGill1} on the BCI data set was based on
computer simulations in which there were difficulties in knowing
when to stop the simulations, i.e. when equilibrium had been
reached. It is unclear whether McGill averaged over an ensemble of
runs, which is essential to obtain repeatable and reliable results
from simulations of stochastic processes because of their inherent
noisiness. However, simulations of the neutral theory are no
longer necessary, and all problems with simulations are moot,
because an analytical solution is now available.

The lognormal distribution is biologically less informative and
mathematically less acceptable as a dynamical null hypothesis for
the distribution of RSA than the neutral theory. The parameters of
the neutral theory or RSA are directly interpretable in terms of
birth and death rates, immigration rates, size of the
metacommunity, and speciation rates. A dynamical model of a
community cannot yield a lognormal distribution with finite
variance because in its time evolution, the variance increases
through time without bound. However, as shown by Sugihara et
al.\cite{Sugihara1}, the lognormal distribution can arise in
static models, such as those based on niche hierarchy.

The steady-state deficit in the number of rare species compared to
that expected under the logseries can also occur because rare
species grow differentially faster than common species and
therefore move up and out of the rarest abundance categories due
to their rare species advantage\cite{Chave1}. Indeed, it is likely
that several different models (e.g. an empirical lognormal
distribution, niche hierarchy models\cite{Sugihara1} or the theory
presented here) might provide comparable fits to the RSA data (we
have found that the lognormal does slightly better than the
neutral theory for the Pasoh data set\cite{Manokaran1}, a tropical
tree community in Malaysia). Such fitting exercises in and of
themselves, however, do not constitute an adequate test of the
underlying theory. Neutral theory predicts that  the degree of
skewing of the RSA distribution ought to increase as the rate of
immigration into the local community decreases. Dynamic data on
rates of birth, death, dispersal and immigration are needed to
evaluate the assumptions of neutral theory and determine the role
played by niche differentiation in the assembly of ecological
communities.

Our analysis should also apply to the field of population genetics
in which the mutation-extinction equilibrium of neutral allele
frequencies at a given locus has been studied for several decades
\cite{Kimura1,Ewens1,Karlin1,Watterson1,Kimura2,Kimura3}.


\bibliography{ecology}

\vspace{1cm}

\noindent {\bf Acknowledgements} We are grateful to Oleg
Kargaltsev for a careful reading of the manuscript. This work was
supported by COFIN MURST 2001, NASA, by NSF IGERT grant
DGE-9987589, by NSF grants DEB-0075102, DEB-0108380, DEB-0129874,
and DEB-0206550 and by the Department of Plant Biology, University
of Georgia.

\vspace{1cm}

\noindent {\bf Competing interests statement} The authors declare
that they has no competing financial interests.

\vspace{1cm}

\noindent {\bf Correspondence} and requests for materials should
be addressed to JRB (banavar@psu.edu) or to SPH
(shubbell@dogwood.botany.uga.edu).

\newpage

{\bf \large Box 1}

{\bf Derivation of the RSA of the local community}

We study the dynamics within a local community following the mathematical framework of McKane et
al.\cite{McKane1}, who studied a mean-field stochastic model for species-rich assembled communities. In our
context, the dynamical rules\cite{Hubbell1} governing the stochastic processes in the community are:

1) With probability $1-m$, pick two individuals at random from the
local community. If they belong to the same species, no action is
taken. Otherwise, with equal probability, replace one of the
individuals with the offspring of the other. In other words, the
two individuals serve as candidates for death and parenthood.

2) With probability $m$, pick one individual at random from the
local community. Replace it by a new individual chosen with a
probability proportional to the abundance of its species in the
metacommunity. This corresponds to the death of the chosen
individual in the local community followed by the arrival of an
immigrant from the metacommunity. Note that the sole mechanism for
replenishing species in the local community is immigration from
the metacommunity, which for the purposes of local community
dynamics is treated as a permanent source pool of species, as in
the theory of island biogeography\cite{MacArthur1}.

These rules are encapsulated in the following expressions for
effective birth and death rates for the $k$-th species:

\begin{equation}
\label{e85} b_{n,k}=(1-m){n\over J}{J-n\over J-1}+m{\mu_k\over J_M}\left(1-{n\over J}\right),
\end{equation}

\begin{equation}
\label{e86} d_{n,k}=(1-m){n\over J}{J-n\over
J-1}+m\left(1-{\mu_k\over J_M}\right){n\over J},
\end{equation}
where $\mu_k$ is the abundance of the $k$-th species in the
metacommunity and $J_M$ is the total population of the
metacommunity.

The right hand side of Eq.(\ref{e85}) consists of two terms. The
first corresponds to Rule (1) with  a birth in the $k$-th species
accompanied by a death elsewhere in the local community. The
second term accounts for an increase of the population of the
$k$-th species due to immigration from the metacommunity. The
immigration is, of course, proportional to the relative abundance
$\mu_k/J_M$ of the $k$-th species in the metacommunity.
Eq.(\ref{e86}) follows in a similar manner. Note that $b_{n,k}$
and $d_{n,k}$ not only depend on the species label $k$ but also
are no longer simply proportional to $n$.

Substituting Eq.(\ref{e85}) and Eq.(\ref{e86}) into Eq.(\ref{e2}),
one obtains the expression\cite{McKane1}

\begin{equation}
\label{e87} P_{n,k}={J!\over
n!(J-n)!}{\Gamma(n+\lambda_k)\over\Gamma(\lambda_k)}{\Gamma(\vartheta_k-n)\over
\Gamma(\vartheta_k-J)}{\Gamma(\lambda_k+\vartheta_k-J)\over\Gamma(\lambda_k+\vartheta_k)}\equiv
F(\mu_k),
\end{equation}
where

\begin{equation}
\label{e88} \lambda_k={m\over (1-m)}(J-1){\mu_k\over J_M}
\end{equation}
and

\begin{equation}
\label{e89} \vartheta_k=J+{m\over (1-m)}(J-1)\left(1-{\mu_k\over J_M}\right).
\end{equation}


Note that the $k$ dependance in Eq.(\ref{e87}) enters only through
$\mu_k$. On substituting Eq.(\ref{e87}) into Eq.(\ref{n1}), one
obtains

\begin{equation}
\label{n891} \langle\phi_n\rangle =\sum_{k=1}^{S_M} F(\mu_k)=
S_M\langle F(\mu_k)\rangle=S_M\int d\mu \widehat{\rho}(\mu)
F(\mu).
\end{equation}
Here $\widehat{\rho}(\mu)d\mu$ is the probability distribution of
the mean populations of the species in the metacommunity and has
the form of the familiar Fisher logseries (in a singularity-free
description\cite{Fisher1,Rao1})

\begin{equation}
\label{e81}
\widehat{\rho}(\mu)d\mu={1\over\Gamma(\varepsilon)\delta^\varepsilon}\exp(-\mu/\delta)\mu^{\varepsilon-1}d\mu,
\end{equation}
where $\delta= {x\over 1-x}$. Substituting Eq.(\ref{e81}) into the
integral in Eq.(\ref{n891}), taking the limits $S_M\rightarrow
\infty$ and $\varepsilon\rightarrow 0$ with $\theta=S_M
\varepsilon$ approaching  a finite value\cite{Fisher1,Rao1} and on
defining $y=\mu {\gamma\over\delta\theta}$, one obtains our
central result Eq.(\ref{e91}).

\newpage


\begin{figure}
\includegraphics[scale = 0.7]{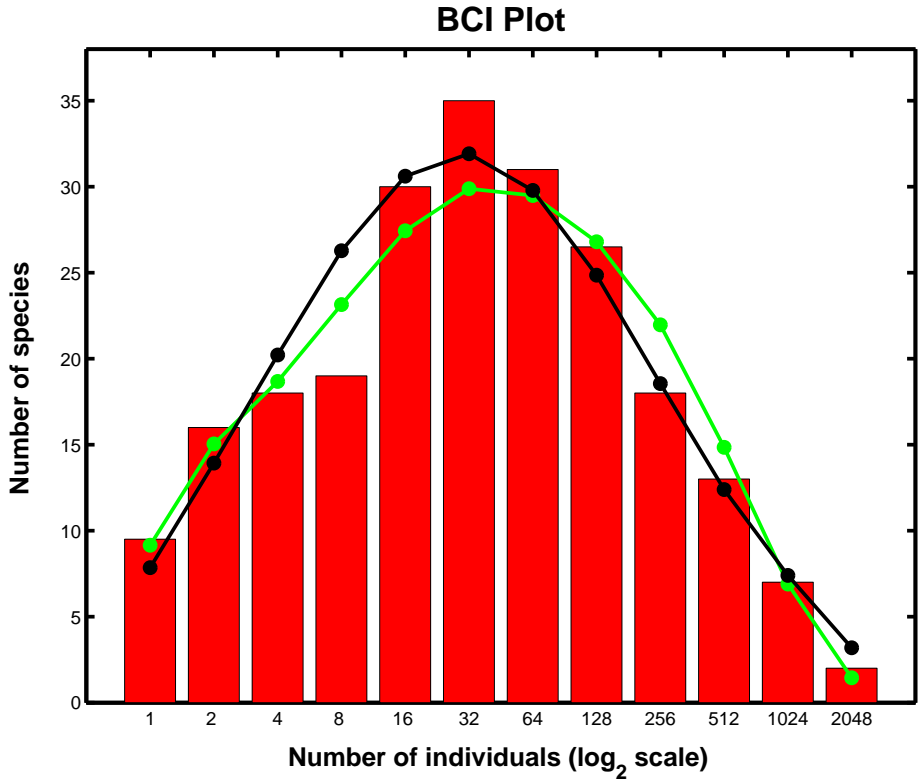}
\caption{\label{fig:condit} Data on tree species abundances in $50$
hectare plot of tropical forest in Barro Colorado Island, Panama
taken from Condit et al.\cite{Condit1}. The total number of trees in
the dataset is $21457$ and the number of distinct species is $225$.
The red bars are observed numbers of species binned into $\log (2)$
abundance categories, following Preston's method\cite{Preston1}. The
first histogram bar represents ${\langle\phi_1\rangle\over 2}$, the
second bar ${\langle\phi_1\rangle\over 2}+{\langle\phi_2\rangle\over
2}$, the third bar ${\langle\phi_2\rangle\over
2}+\langle\phi_3\rangle+{\langle\phi_4\rangle\over 2}$, the fourth
bar ${\langle\phi_4\rangle\over
2}+\langle\phi_5\rangle+\langle\phi_6\rangle+\langle\phi_7\rangle+{\langle\phi_8\rangle\over
2}$ and so on. The black curve shows the best fit to a lognormal
distribution $\langle\phi_n\rangle={N\over n}\exp(-{(\log_2 n-\log_2
n_0)^2\over 2\sigma^2})$ ($N=46.29$, $n_0=20.82$ and $\sigma=2.98$),
while the green curve is the best fit to our analytic expression
Eq.(\ref{e91}) ($m=0.1$ from which one obtains $\theta=47.226$
compared to the Hubbell\cite{Hubbell1} estimates of $0.1$ and $50$
respectively and McGill's best fits\cite{McGill1} of $0.079$ and
$48.5$ respectively.)}
\end{figure}

\end{document}